# The Role of Patents: Incentivizing Innovation or Hindering Progress?

Gaétan de Rassenfosse

Holder of the Chair of Science, Technology, and Innovation Policy
École polytechnique fédérale de Lausanne, Switzerland.

**Purpose**
This article is part of a Living Literature Review exploring topics related to intellectual property, focusing on insights from the economic literature. Our aim is to provide a clear and non-technical introduction to patent rights, making them accessible to graduate students, legal scholars and practitioners, policymakers, and anyone curious about the subject.

**Funding**
This project is made possible through a Living Literature Review grant generously provided by Open Philanthropy. Open Philanthropy does not exert editorial control over this work, and the views expressed here do not necessarily reflect those of Open Philanthropy.

## The Role of Patents: Incentivizing Innovation or Hindering Progress?

Gaétan de Rassenfosse
École polytechnique fédérale de Lausanne, Switzerland.

Simply put, a patent is a legal document securing the right to exclude others from making, using, or selling an invention for several years. At first glance, the idea of granting exclusive rights to an invention might seem surprising—perhaps even unsettling—since we often think of knowledge as a shared resource. Wouldn't the world benefit more if knowledge flowed freely for everyone to use?

While it is hard to argue against the idea that knowledge should be shared as widely as possible, it is important to ensure that new knowledge is being created in the first place. That's where patent rights come into play, as we explore below.

Before diving into why the patent system exists, however, it is worth noting that most inventions are *not* subject to patent protection. The United States Patent and Trademark Office (USPTO) issued its twelfth millionth patent in mid-2024. Of these, around five million patents were granted in the last two decades and could still be in force. However, patents do not automatically stay valid for their full 20-year term—owners must pay renewal fees regularly to maintain them. Many choose not to, so a rough estimate is that only about three million inventions are currently protected by active patents. Furthermore, not all patentable inventions are patented—many are kept secret—and scientific discoveries are not eligible for patenting. According to OpenAlex, a comprehensive open dataset of scholarly work, approximately 120 million scientific articles have been produced over the past two decades—far exceeding the number of patents granted in the same period. In other words, only a tiny portion of the world's knowledge is protected by patents.

### What is all the fuss about patents, then?

If only a small fraction of knowledge is protected by patents, why are patents such a polarizing topic? One reason lies in the philosophical debate surrounding the nature of knowledge itself. Many believe that knowledge, as a universal and shared resource, transcends individual ownership and should remain free from barriers, serving the collective progress of humanity.

Beyond this philosophical divide, patents stir controversy because of their real-world implications. Critics argue that patents limit the use of knowledge, particularly in critical fields like medicine, where restricted access to patented drugs can have life-or-death consequences. The debate surrounding patent protection for antiretroviral medications during the AIDS crisis at the turn of the millennium is a stark example of these tensions. Patent protection helped keep prices high, effectively excluding patients in the least-developed countries from accessing life-saving treatments. More recently, similar debates resurfaced with COVID-19 vaccines, as many stakeholders advocated for waiving patent rights to ensure equitable global access.

These arguments, and others, are used by some to argue that the patent system should be abolished, pure and simple. However, economists studying the patent system generally hold a more nuanced view, as explained below.

**Why do we have a patent system after all?**

The answer to that question often lies in the economic rationales offered to justify the patent system. However, these rationales are ex-post explanations constructed long after the system's inception. In reality, we have a patent system today largely because, on March 19, 1474, the Senate of Venice happened to enact the Venetian Patent Statute—a legal innovation shaped by the unique political and economic conditions of Renaissance Venice. The Statute recognized that creative minds thrived in Venice's bustling and innovative environment, and it sought to incentivize these inventors by granting them exclusive rights to their creations for up to ten years. This protection ensured that others could not copy their work and "take their honor," providing a strong motivation to invent and contribute to the state's prosperity.

What began as a local measure to encourage invention became the foundation of a global system. Over the centuries, governments across Europe and beyond saw the potential of such systems to foster technological progress and economic development, leading to the evolution of the patent systems we use today. In the United States, the importance of patents was enshrined in the Constitution, which explicitly grants Congress the power to promote the progress of science and useful arts by issuing exclusive rights. From the 15th-century canals of Venice to the globalized innovation landscape of the 21st century, the patent system has evolved through countless amendments and adjustments, shaped by the changing needs of societies and economies.

The understanding of the patent system has evolved alongside advances in theory and analytical tools available to economists. Menell (1999) provides an excellent overview of the evolution of the so-called utilitarian theory of patent rights, beginning with the influential political economists of the 18th and 19th centuries. Pioneers like Adam Smith, Jeremy Bentham, and John Stuart Mill were among the first economists to argue for granting inventors limited "monopolies" to promote innovation. Building upon these insights, prominent economists such as Arthur Cecil Pigou, John Bates Clark, and Kenneth Arrow further refined economic thinking about intellectual property. Their collective work culminated in what is now considered the classical framework for modeling the patent system, formalized by Nordhaus (1969).

William Nordhaus' model focuses on the optimal duration of patent protection, recognizing that a longer patent term incentivizes innovation by letting inventors earn more profit but also harms society by extending the period of monopoly pricing. The optimal patent length, according to Nordhaus, balances these competing forces. The Nordhaus model highlights the fundamental trade-off in patent policy. On the one hand, longer patent terms provide greater incentives for innovation by allowing inventors to capture more of the social benefits through higher profits. On the other hand, longer patent terms also prolong the period of monopoly pricing, leading to reduced consumer welfare.

**Why do we need to incentivize innovation?**

The theory behind the patent system rests on the idea that we need to incentivize innovation. But does this mean society would cease inventing if the patent system were abolished? The answer is a resounding no. Ingenuity and creativity are fundamental aspects of human nature, and people would undoubtedly continue to invent. In fact, we already know that only a small fraction of inventions are ever patented, which suggests that many inventions emerge and thrive without the promise of exclusive rights. Inventions would likely continue to see the light of day, driven by inventors' curiosity, necessity, and desire to solve problems.

However, the question is not whether society would stop inventing but rather how many inventions society would create without a patent system. A straightforward argument is that patents, by granting exclusive rights to inventors, make it harder for others to build on patented inventions, thereby *limiting* follow-on innovation. In this view, abolishing patent rights could potentially free millions of inventions for society to use and improve. However, this reasoning overlooks a critical point: many of these inventions were created in the first place because of the incentives provided by the patent system. Economists caution that, without the patent system, society might not innovate at the same pace or scale as it does today.

To understand why this is the case, we must introduce the concept of knowledge as a public good. Samuelson (1954: 387) provides the classical definition of a public good as one that "all enjoy in common[,] in the sense that each individual's consumption of such a good leads to no subtraction from any other individual's consumption of that good." Public goods have two key traits: non-excludability and non-rivalry. Non-excludability means that others cannot be prevented from benefiting from the good once it is produced, whereas non-rivalry means that one person's use of the good does not diminish its availability to others. Knowledge is the quintessential example of a public good. Consider the Pythagorean theorem: "In a right triangle, the square of the hypotenuse is equal to the sum of the squares of the other two sides." One architect's use of the theorem does not reduce its availability for others, nor can anyone be prevented from using it.

Recognizing the public-good nature of knowledge, Arrow (1962) concluded that a free market economy often underinvests in innovation compared to the social optimum. From a firm's perspective, investment decisions in R&D are driven by expected private profits, which fail to account for broader societal benefits, such as consumer surplus. Moreover, the non-excludability of knowledge means that innovators cannot fully capture the benefits of their creations, as knowledge can diffuse to competitors and other firms. Consequently, projects with high social value but insufficient private returns are frequently overlooked, depriving society of innovations that could have significantly improved collective welfare.

The gap between social and private returns, and the resulting underinvestment in R&D by the free market, is a primary justification for R&D support mechanisms such as grants and subsidies. The patent system is another such mechanism, specifically designed to strengthen the appropriability of inventions. By allowing inventors to exclude others from using their inventions, patents increase the private returns on R&D investments, thereby fostering greater overall R&D activity in the economy. The exclusivity conferred by patents is not a flaw but an intentional feature of the system. While patent protection enables firms to charge temporarily higher prices for their innovative products—limiting their immediate diffusion in society—this trade-off represents the (imperfect) solution to incentivize firms to invest in developing these products in the first place. Economies accept the 'static inefficiency' of higher prices because the 'dynamic efficiency'—greater incentives to invest in innovation—presumably outweighs it.

**From the theory of patents to the design of the patent system**

The theoretical considerations thus far allow us to understand some important features of the patent system. Following the theory's logic, patents would only be granted for inventions that would not have been created if the patent system did not exist. Granting exclusivity to inventions that would have been developed anyway creates monopolies without delivering the intended incentive effect. However, it is exceedingly difficult for patent offices to determine whether an invention was motivated by the prospect of higher profits delivered by the patent system. In practice, patent offices grant protection based on criteria such as worldwide novelty, non-obviousness, and utility. Worldwide novelty ensures that the invention is genuinely new, while the non-obviousness requirement—indicating an inventive step—can be seen as a proxy for identifying inventions that are sufficiently innovative to justify exclusivity.

From an economic efficiency perspective, knowledge should be freely distributed to ensure its optimal utilization. This is because the marginal cost of providing existing knowledge to an additional person is effectively zero, reflecting the non-rivalry aspect of knowledge. This principle aligns with the concept of marginal-cost pricing, a cornerstone of welfare economics, which dictates that goods and services should be priced at their marginal cost of production to achieve allocative efficiency (Arrow 1962). While patents limit the use of an invention by others, the patent document provides a detailed description of the invention, freely accessibly to everyone. Patent collections thus serve as publicly available repositories of knowledge from which inventors can learn.

The requirement for patent documents to disclose the details of inventions serves several practical purposes: enabling examiners to assess novelty, allowing inventors to clearly define their claims, and helping others avoid infringing on the protected technology. Beyond these practicalities, the temporary exclusivity granted in exchange for public disclosure is often described as a "bargain" (Hall, 2007) or a "social contract" (Levin, 1986) between inventors and society, or the "quid pro quo" of the patent system (Machlup and Penrose, 1950). Moreover, patent protection is both time-bound and geographically limited. Patents are jurisdictional rights, valid only in the specific country where protection is sought and granted. Obtaining worldwide protection is prohibitively expensive, so most inventors seek patents in only a select few countries. As a result, most inventions are freely available for use in many parts of the world.

The disclosure requirement, along with the temporal and geographic limitations of patent protection, reduces the scope of the exclusion rights conferred by patents, partially mitigating the economic inefficiencies associated with restricted utilization. However, some scholars have argued that patents can confer exclusion rights that are excessively strong, potentially rendering the patent system inefficient. Boldrin and Levine (2002) suggest that the current intellectual property system, by granting excessive control over the use of ideas, often creates more problems than it resolves. They distinguish between two components of patent rights: the right to own and sell ideas—commonly referred to as the right of first sale—and the right to control the use of those ideas after the sale, known as the downstream licensing right. While they see the first component as essential, they view the second as economically harmful. They argue that allowing producers to dictate how consumers use purchased products containing intellectual property stifles competition from those consumers, weakening market competition overall.

Ultimately, the necessity of the patent system is for readers to judge. Perhaps the best way to conclude is with the words of economist Edith Penrose (1951: 40), who aptly observed: “If national patent laws did not exist, it would be difficult to make a conclusive case for introducing them; but the fact that they do exist shifts the burden of proof and it is equally difficult to make a really conclusive case for abolishing them.”